\documentclass[fleqn,10pt]{olplainarticle}

\usepackage{timet}

\usepackage{amsmath}
\usepackage{amsfonts}
\usepackage{graphicx}
\graphicspath{{figs/}}
\usepackage{physics}
\usepackage[normalem]{ulem}
\usepackage{cancel}
\usepackage{xcolor}

\usepackage{tikz}
\usepackage{url}

\tikzset{>=latex}

\title{Quantitative imaging of the fresh/saltwater interface with airborne electromagnetics: examining different sources of uncertainty}

\author[Wouter Deleersnyder]
{Wouter Deleersnyder$^{1,2}$,  David Dudal$^{1}$, Thomas Hermans$^{2}$\\
	$^{1}$KU Leuven Campus Kortrijk - KULAK, Department of Physics, Etienne Sabbelaan 53, 8500 Kortrijk, Belgium.\\ 
	$^{2}$Ghent University, Department of Geology, Krijgslaan 281 - S8, 9000 Gent, Belgium \\  E-mail: {wouter.deleersnyder@ugent.be} \\
}
\keywords{Uncertainty Quantification, Salinity, Airborne, Electromagnetics}

\begin{abstract}
Knowing the distribution between fresh and saline groundwater is imperative for sustainable and integrated management of water resources in coastal areas. The airborne electromagnetic (AEM) method is increasingly used for hydrogeological mapping over large areas via bulk electrical resistivity. However, accurately and reliably mapping the fresh/saltwater interface (FSI) requires accurate knowledge about the transition zone. The objective is to quantify the uncertainty in using AEM data to inform on the depth of the FSI. The study mimics a dual-moment time-domain SkyTEM sounding recorded in the Belgian coastal plain based on borehole data. It quantifies uncertainty using a differential evolution adaptive Metropolis algorithm to sample the posterior distribution. The results indicate the importance of reliable altitude, pitch and roll logging. Gathering prior knowledge about the transition zone, for example, through borehole logs, significantly improves the estimation of the FSI. The Resolve frequency-domain system, especially in context with very shallow to shallow FSIs, is more suitable for salinity mapping than the time-domain SkyTEM used in the field survey. The depth of the FSI may be defined via various threshold values. The uncertainty of three different thresholds is studied. The FSI based on the middle of the transition zone is the most reliable, while the FSI based on the 1500 mg/L total dissolved solids threshold is the least robust.
\end{abstract}

\begin{document}

\flushbottom

\maketitle
\thispagestyle{empty}

\section{Introduction}
Planning for water security in the context of global environmental change is one of the most pressing challenges worldwide, as articulated by the 6th Sustainable Development Goal: Clean water and sanitation \citep{UNESCO_2019, taylor2013ground, elshall2020groundwater}. Knowing the distribution between fresh and saline groundwater is imperative for sustainable and integrated management of water resources in coastal areas. The Belgian coastal plain is quite a complex distribution of fresh and saltwater. The distribution is a result of the geologic evolution after the last ice age (salinization) and recent human intervention by land reclamation and impoldering (see, e.g. \citet{vandenbohede2012groundwater, vandenbohede2014quartair}). The latter caused freshwater lenses on top of ‘fossil’ saltwater. Precipitation deficits due to the drier summers or overexploitation of these freshwater lenses can (negatively) affect the saltwater interface. Moreover, rising sea levels will put pressure on the freshwater lenses in the dunes, which could increase saltwater intrusion and the polders’ saline seepage pressure. This effect could negatively affect the future underground bearing of precipitation surpluses \citep{oude2010effects}.Therefore, Flanders Environmental Agency (VMM) wanted to update an older freshwater/saltwater interface map \citep{debreuck1989} to prepare for future climate adaptation. From 2017 to 2019, a SkyTEM airborne electromagnetic survey covering the coastal area was obtained, which was commissioned by VMM. The survey recorded over 67,500 soundings along 2412 km with a 250 m line spacing \citep{vlaanderentopsoil}. \\

Airborne Electromagnetic ({AEM}) mapping, in general, is an increasingly used method to image near-surface geological features over large areas via bulk electrical resistivity. Common applications are, for example, in mineral exploration \citep{macnae2007developments}, contamination \citep{pfaffhuber2017delineating}, hydrogeological mapping \citep{mikucki2015deep}, and saltwater intrusion \citep{pedersen2017mapping,siemon2019automatic,gottschalk2020using, deleersnyder2023flexible}. While for some applications, anomalies in the data be directly interpreted, other applications require geophysical inversion which seeks inversion models matching the observed data. For some saltwater imaging applications, when one is only interested in the qualitative distribution of the fresh and saltwater lenses, standard inversion approaches may be sufficient \citep{viezzoli2009spatially}, but mapping the fresh/saltwater interface (FSI) requires an accurate knowledge about the transition from fresh to salt. Otherwise, one may find a deviation of a few meters. However, \citet{king2022airborne} and \citet{deleersnyder2023improving} demonstrate that accurately imaging the electrical conductivity right on top of a saltwater lens may yield highly uncertain results, impeding such quantitative mapping.\\

This uncertainty is related to the non-unicity of the inverse problem. Multiple inversion models fit the data equally well. Standard interpretation approaches rely on deterministic inversion, in which the goal is to locate one optimal model that can be obtained using some form of smoothness or sharpness constraints implied through several available regularization techniques \citep{constable1987occam, farquharson1998non,vignoli2015sharp, viezzoli2009spatially, deleersnyder2021inversion}. There is, however, no guarantee that the specific regularization scheme will represent realistic features. For example, with Tikhonov regularization, overly smooth inversion models can be obtained, which will impact the reliability of the depth of such a FSI (In Section \ref{subsec:thresholds}, the various thresholds in Electrical Conductivity (EC) at which the FSI is defined are described).\\

Probabilistic methods offer an alternative in which the solution is not one model, but a collection of thousands of models all of which fit the data within the noise level. From the ensemble, statistical properties about the non-uniqueness and uncertainty of the models can be derived. Probabilistic inversion has already been conducted for AEM data \citep{hansen2019inversion, hansen2021efficient, bai20211d} and with SkyTEM data \citep{brodie2013monte}. To our knowledge, the uncertainty related to the estimation of the depth of an interface in a fresh/saltwater environment with EM has not yet been studied. Moreover, we will test the reliability of three different definitions of such a threshold.\\

Faster (approximate) alternatives exist, such as Bayesian Evidential Learning (BEL) \citep{ahmed2024assessing} or deep learning approaches \citep{hansen2022use,oh2021bayesian}. Those alternatives exist because probabilistic inversion for highly-dimensional problems tends to be computationally very demanding, which is why deterministic approaches prevail today. Let us mention that alternative deterministic inversion schemes exist, which can tune the degree of sharpness \citep{klose2022laterally, deleersnyder2023flexible} and, therefore, can generate an ensemble of inversion models with different features. In \citet{deleersnyder2024validation}, we have found that such a wavelet-based inversion scheme can create an ensemble with a wide range overlapping with the true posterior, making them useful to qualitatively assess the range of the uncertainty. However, it tends to sample shallower interfaces more often, which is attributed to the minimum-structure inversion. The density of the samples can thus not be interpreted as a distribution, and full probabilistic approaches are still needed for an accurate uncertainty quantification. Here, we have sufficient prior knowledge and we will limit the computational burden of the probabilistic approach in that way. \\

This work aims to quantify the uncertainty using AEM data to inform the FSI depth. It studies the effect of flight altitude, altitude uncertainty, and tilt of the transmitter-receiver frame, compares it with an FDEM system, and examines the effect of prior knowledge about the sharpness and salinity of the saltwater. Following the \citet{hansen2019inversion} philosophy, we explicitly choose a prior model based on the available information (borehole information); see Section \ref{subsec:parameterization}. A realistic synthetic model is constructed, and corresponding data is generated to exactly mimic field conditions according to the 2019 SkyTEM survey  \citep{vlaanderentopsoil}. Our probabilistic inversion scheme, explained in Section \ref{subsec:probabilistic}, relies on the Differential evolution Markov chain by \citet{laloy2012high}. Section \ref{subsec:thresholds} discusses the different possibilities for thresholds for a FSI. After the results in Section \ref{sec:results}, some suggestions for further research and new surveys with a view to mapping the FSI are listed in Section \ref{sec:conclusion}.

\section{Methods}
\label{sec:methods}
\subsection{Probabilistic inversion}
\label{subsec:probabilistic}
\citet{tarantola1982generalized} developed a Bayesian approach to solving inverse problems, in which all information has be to be quantified probabilistically. A prior distribution represents the prior knowledge, from geological expert knowledge or previous surveys. Secondly, the likelihood describes how well the forward response matches the observed data:
\begin{equation}
	L(\vb{m}) \sim \prod_{i}\exp\left( -\frac{1}{2} \frac{(d_{i, \text{obs}} - \mathcal{F}_i(\vb{m}) )^2}{\sigma_i^2} \right), 
\end{equation}
where $\vb{m}$ represents the inversion model and $\mathcal{F}(\vb{m})$ describes the forward relation, a semi-analytical solution for the magnetic field response for a 1D (horizontally) layered earth by \citet{hunziker2015electromagnetic, werthmueller2017open}. $\sigma$ represent the estimated noise, which is a combination of a multiplicative noise (here 3\% of the observed data) and a noise floor ($10^{-12}$, realistic estimate for the SkyTEM system \citep{auken2009integrated}). Via the so-called conjunction of states of information, both the prior distribution and the likelihood are combined via multiplication. For general nonlinear cases, one must resort to sampling techniques to generate a set of models $\vb{m}$ that are proportional to the true posterior distribution. \\

To sample the posterior, we use the algorithm presented by \citet{laloy2012high},  using multiple-try differential evolution adaptive Metropolis (DREAM$_\text{ZS}$), which combines the strengths of multiple-try sampling, snooker updating, and sampling from an archive of past states. 

\subsection{Parametrization}
\label{subsec:parameterization}
\begin{figure}
	\centering
	\includegraphics[width=0.495\linewidth]{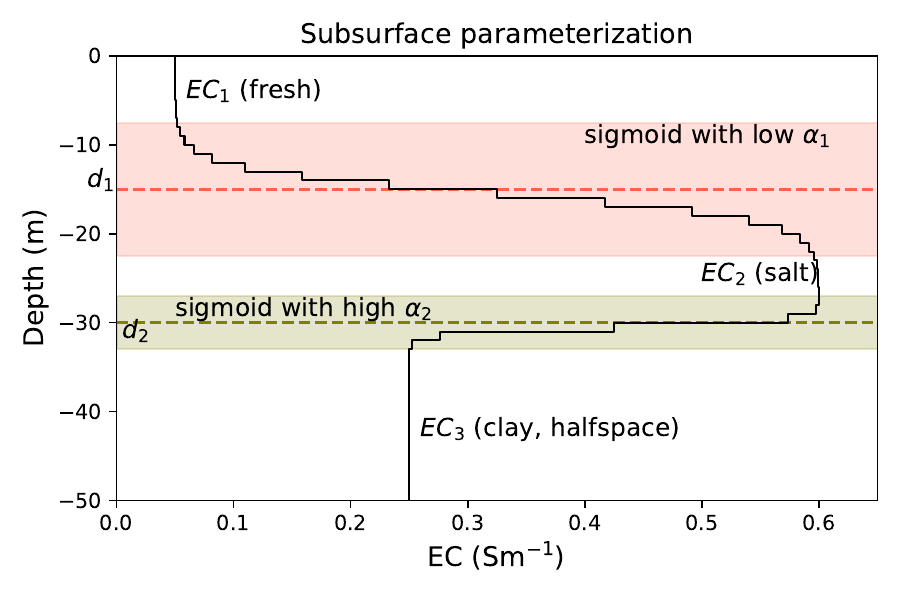}
	\caption{The parametrization \eqref{eq:parameterization} consists of a freshwater lens on top of a saltwater lens resting on a clay formation. Each transition is modelled with a logistic sigmoid with a sharpness parameter $\alpha_i$ at a depth $d_i$.}
	\label{fig:parametrization}
\end{figure}

We rely on implicit model assumptions to limit the computational burden, representing prior information from the study area \citep{hansen2019inversion}. Based on borehole loggings, we can parameterize the subsurface according to a typical shape (see Figure \ref{fig:parametrization}) for the study area: a mainly sandy sediment layer with fresh water, a sandy sediment layer with saltwater, a halfspace consisting of clay, and transition zones in between the layers with a variable sharpness. We model the transition with a logistic sigmoid function. The model parameterization consists out of only 7 parameters and is as follows:
\begin{equation}
	\label{eq:parameterization}
	EC(y) = EC_1 + \frac{EC_2 - EC_1}{1 + \exp\left(-\alpha_1  (y - d_1)\right)} - \frac{EC_2 - EC_3}{
		1 + \exp\left(-\alpha_2 (y - d_2)\right)},
\end{equation}
where $EC_1$, $EC_2$ and $EC_3$ are the electrical conductivity of the bulk with freshwater, saltwater and clayey halfspace respectively. Each transition has a sharpness parameter $\alpha_i$ and a depth $d_i$. An alternative would be to consider spatially correlated Guassian models to reduce the complexity \citep{hansen2009reducing}, especially if such an explicit parameterization cannot be derived from prior knowledge. The advantage of having an explicit sharpness parameter is that we can easily test the effect of having more/less prior knowledge about the transition zone, see Section \ref{subsec:sharpness}.\\

Actual borehole data displays more variability than the proposed parameterization in \eqref{eq:parameterization}. Borehole EM logs, however, have a higher vertical resolution, and AEM may not capture the observed variability making this parameterization a realistic and effective way to reduce the dimension of the inverse problem. Moreover, AEM has a large footprint, which averages out such variations. 

\subsection{Deriving the depth of the interface}
\label{subsec:thresholds}
The TDEM method is mainly sensitive to bulk electrical conductivity, which depends on the soil properties (porosity, clay content) and the conductivity of the groundwater in the pore space. A petrophysical law, such as Archie's law \citep{archie1942electrical}, links the bulk conductivity to the conductivity of the water. Cases with mainly sandy sediments above the clayey halfspace are prevailing and are for simplicity considered here.\\

The FSI in the study area was defined at a 1500 mg/L TDS limit, which, via a statistical analysis of water samples, is linked to a water conductivity of 0.2 Sm$^{-1}$. Via the petrophysical law applied in sandy sediments, a bulk conductivity of 0.05~Sm$^{-1}$ is obtained. With deterministic inversion, however, the non-uniqueness right on top of a highly conductive lens is significant, and the traditional smoothness constraint yields slowly increasing conductivity models, which reach the 0.05~Sm$^{-1}$  threshold too early, yielding a too-conservative estimation of the depth of the interface \citep{vlaanderentopsoil}. Hence, this threshold was named the conservative threshold. Another threshold, from brackish to saltwater, occurring at 0.5 Sm$^{-1}$ water conductivity (or 0.125~Sm$^{-1}$ bulk conductivity in sandy sediments) yielded depth estimations, which corresponded better with the FSIs from the available water samples in \citet{vlaanderentopsoil}. Therefore, this threshold was used as the optimistic threshold to image the FSI.
We introduce a third, alternative threshold to investigate whether defining a threshold less prone to the non-uniqueness of the inverse problem is possible and which will be more reliably recoverable with more traditional deterministic models. We define the alternative threshold at the inflexion point $d_i$ of the transition zone. In Section \ref{subsec:salinity}, the performance of each threshold is tested for environments with different salinities. 

\section{Results}
\label{sec:results}
\begin{figure}
	\centering
	\includegraphics[width=0.495\linewidth]{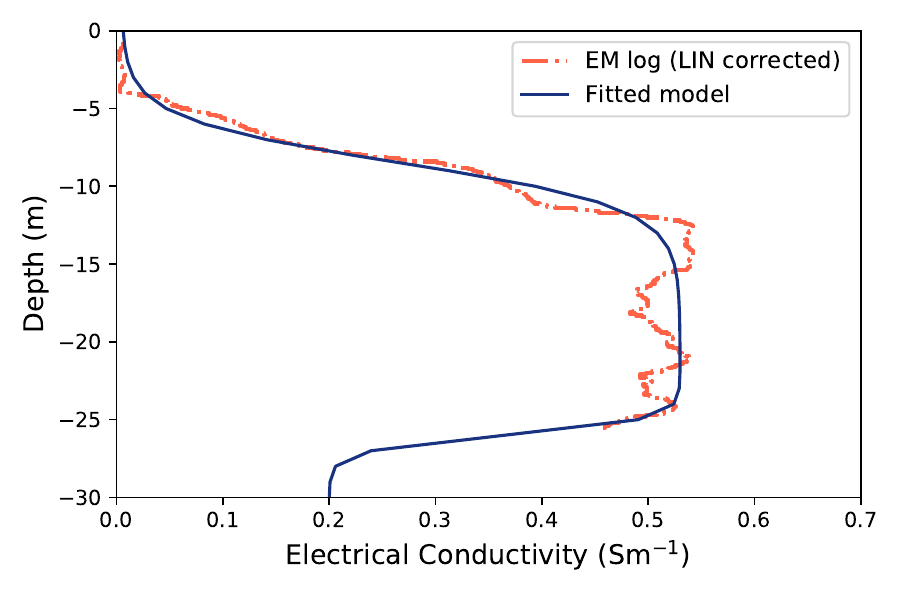}
	\caption{The parameters from Eq. \eqref{eq:parameterization} are fitted from a typical EM borehole log from the study area. The result is the true model, summarized in Table \ref{tab:prior_benchmark}.}
	\label{fig:borehole}
\end{figure}

We work with synthetic data from a realistic true model, summarized in Table \ref{tab:prior_benchmark}. We fit the model parameters of the specific parameterization \eqref{eq:parameterization} to an EM borehole log, shown in Figure \ref{fig:borehole}. We obtain the conductivities 0.005 Sm$^{-1}$, 0.53 Sm$^{-1}$, and 0.2 Sm$^{-1}$ for the freshwater, saltwater, and halfspace, respectively. The transition from fresh to saltwater centres at 8.5 m with a sharpness of 0.7 m$^{-1}$, while the transition from salt to clay is even sharper, namely 2 m$^{-1}$ at 25 m depth. The synthetic data is collected at a height of 30 m, to which 3\% multiplicative Gaussian noise is added.\\

For the synthetic model, we first investigate the effect of the non-unicity of the inverse problem for this type of data in this highly conductive/saline environment on the depth of the interface from fresh to saltwater. We consider a realistic measurement noise level (3\%) and prior knowledge. Then, the effect of the altitude and the uncertainty of the altitude are examined, followed by the effect of pitch and roll of the source-receiver frame/bird. Because the study area exhibits a significant heterogeneity of the saltwater's salinity content/electrical conductivity, we also analyse the effect of the EC of the saltwater lens. The effect of (limited) knowledge about the sharpness of the transition is also investigated. Finally, we examine whether an FDEM system would be more suitable for imaging the depth of the fresh-to-saltwater interface in this context.

\subsection{Inherent non-unicity of the inverse problem}
\label{subsec:benchmark}
This work aims to gain insight into the extent to which the data can inform about the parameters and their uncertainty. Therefore, a broad but realistic prior has been constructed; see Table \ref{tab:prior_benchmark}. The centre is always taken similarly to the true model, with a realistic variation from the true for each parameter. Except for the conductivity of freshwater, there is a significant difference between the true and mean of the prior distribution to sample a sufficiently wide realistic range of ECs. Truncated normal distributions avoid negative electrical conductivities (EC) and depths. The prior will be the starting point for each case; variations will be stated explicitly.\\

\begin{figure}
	\centering
	\includegraphics[width=\linewidth]{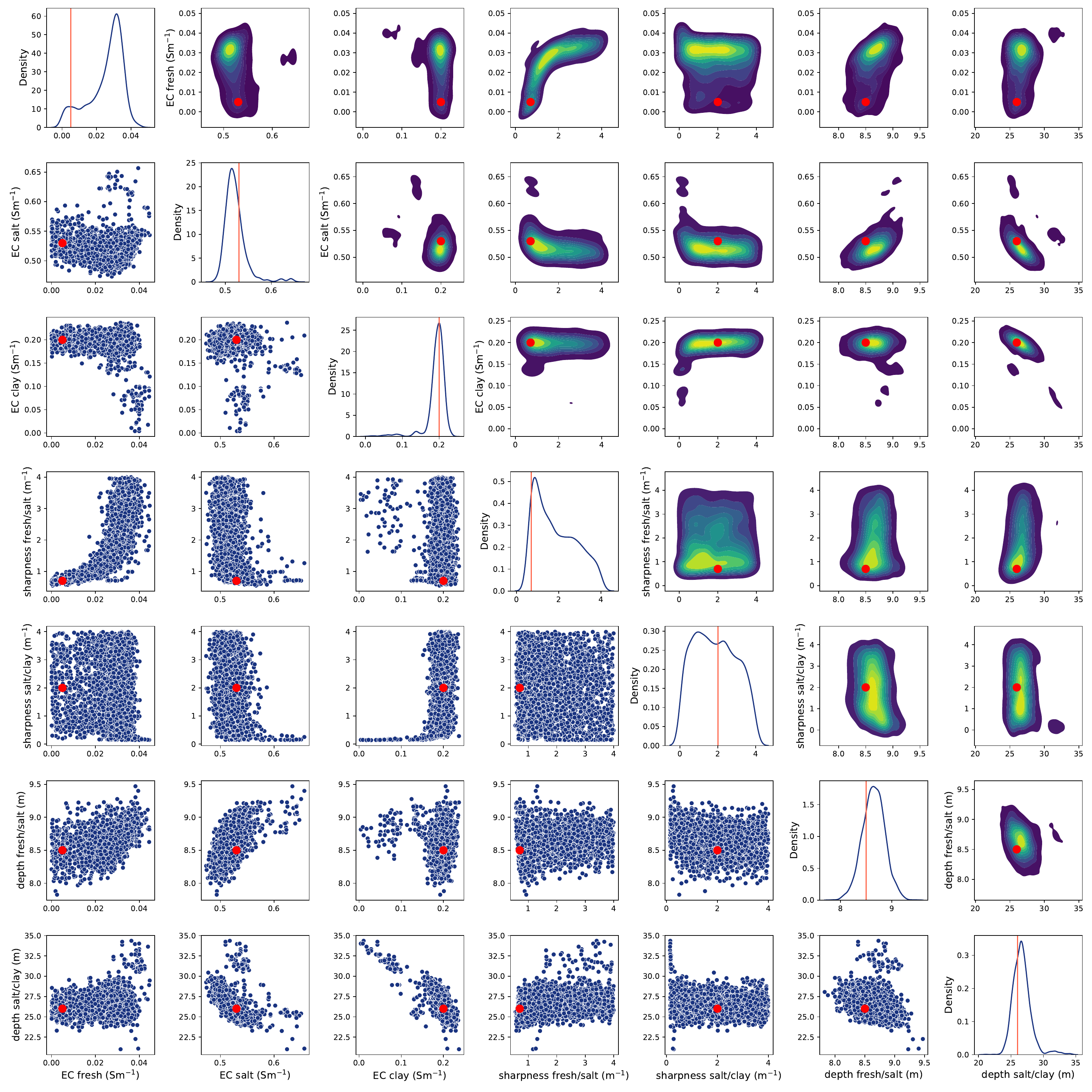}
	\caption{The posterior distribution. Case from Table \ref{tab:prior_benchmark} without altitude or inclination uncertainty. The red dot/vertical lines indicate the true model. }
	\label{fig:benchmark_correlations}
\end{figure}

In this case, the inverse problem's non-unicity is studied, while perfect knowledge of the altitude is assumed. Only models after a burn-in phase are shown, which thus fit the data sufficiently. The posterior distribution is shown in Figure \ref{fig:benchmark_correlations}. The red point/vertical line represents the true model. Through the prior information and observed data interplay, a posterior is found where the maximum corresponds to the true model. Only the EC of freshwater is clearly bimodal, where the maximum does not correspond to the true conductivity of 0.005 Sm$^{-1}$. The centre of the prior yielded a bias towards higher conductivities, but note that the centre of the prior distribution (0.05 Sm$^{-1}$) has a vanishing prevalence. The second local maximum corresponds well with the true conductivity, indicating that the Metropolis-Hastings algorithm could locate this area with a large likelihood based on the observed data, but due to the higher mean of the prior, the absolute maximum of the posterior is in between. Note that based on the representation of the posterior in Figure \ref{fig:benchmark_correlations}, one cannot exclude the occurrence of electrical conductivities of the freshwater larger than 0.05 Sm$^{-1}$; this is an interaction with the other parameters, and the model should be generated based on the parameterization in Eq. \eqref{eq:parameterization}.\\
Moreover, large ECs occur for models with a vanishing conductivity of the clay at large depths together with a very smooth transition from salt to clay. This combination of parameters in the models no longer resembles the structure imposed via parameterization \eqref{eq:parameterization}, and thus no longer accepts the implicitly assumed prior knowledge. This demonstrates the inverse problem's inherent non-unicity and that without prior knowledge (imposed via the prior distribution and/or the parameterization), a much richer set of models could be found that fit the data, hindering further processing.\\

\begin{table}
	\centering
	\begin{tabular}{l|c||c|c|c}
		& True & Mean  & Variance &  Truncation interval \\
		\hline
		$EC_1$ fresh& 0.005 & 0.05  & 0.05   & [0, 0.5] \\
		\hline
		$EC_2$ salt & 0.53  & 0.5  & 0.5  & [0, 1.5]  \\
		\hline
		$EC_3$ clay& 0.2 & 0.25 & 0.2  & [0, 1]   \\
		\hline
		$\alpha_1$ fresh/salt & 0.7 &  1 & 3 & [0, 4]   \\
		\hline
		$\alpha_2$ salt/clay & 2 & 1  & 3  & [0,4 ~]   \\
		\hline
		$d_1$ fresh/salt &  8.5 &  8.5& 5  & [0, 25]  \\
		\hline
		$d_2$ salt/clay & 26 & 25  &  5 & [5, 50]   \\

	\end{tabular}
	\caption{Prior distribution according to parametrization \eqref{eq:parameterization} reflecting the available prior information for the case in Section \ref{subsec:benchmark} without altitude or inclination uncertainty.}
	\label{tab:prior_benchmark}
\end{table}

\begin{figure}
	\centering
	\includegraphics[width=0.495\linewidth]{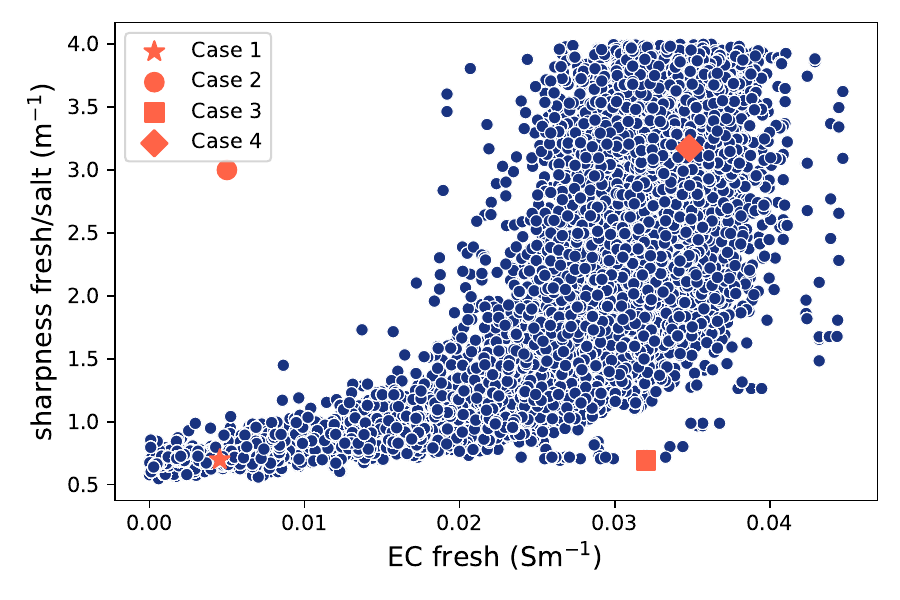}
	\includegraphics[width=0.495\linewidth]{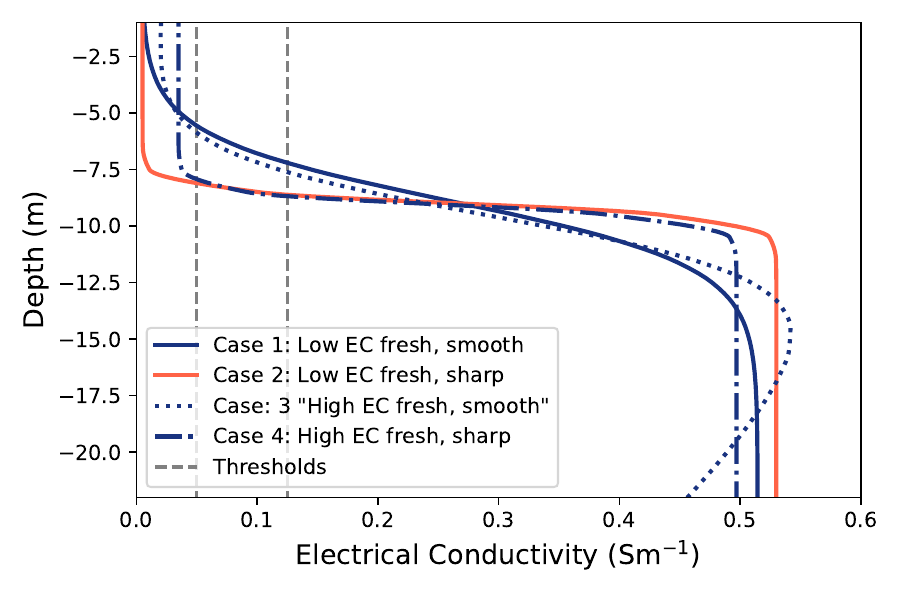}
	\caption{The posterior effectively consists of only two model types: Case 1-like models with low $EC_1$ of the freshwater and smooth transitions and Case 4-like models with relatively high $EC_1$ and a sharp transition.}
	\label{fig:benchmark_interpretation_of_covariance}
\end{figure}

Solely based on the density plots of the two sharpnesses, the observed data does not provide much information. All values between the truncation limits of the prior (see Table \ref{tab:prior_benchmark}) are possible for the transition from salt to clay. Many values are possible for the transition from fresh to salt but with an unambiguous maximum near the true sharpness. The correlation between the sharpness parameter and the freshwater lens's EC in Figure \ref{fig:benchmark_interpretation_of_covariance} provides some insight. Consider four cases with high/low EC and smooth/sharp interfaces. Case 1 is the true model, clearly within the posterior distribution and with low electrical conductivity and sharpness. Case 2 corresponds to a low EC and high sharpness; such cases do not occur in the posterior distribution (hence the red colour). However, a similar case, Case 4, with a sharp transition, fits the data with a larger electrical conductivity. Case 3 is from a less densely populated region of the posterior space. Based on the figure on the left, you would classify this case as a smooth model with high electrical conductivity. However, the sharpness of the second transition from salt to clay causes the actual EC of the freshwater lens to be pushed towards lower conductivities. Thus, there are effectively only two regimes: models with smooth transitions and low ECs or sharp transitions and relatively high ECs of the freshwater lens.\\

\begin{figure}
	\centering
	\includegraphics[width=0.495\linewidth]{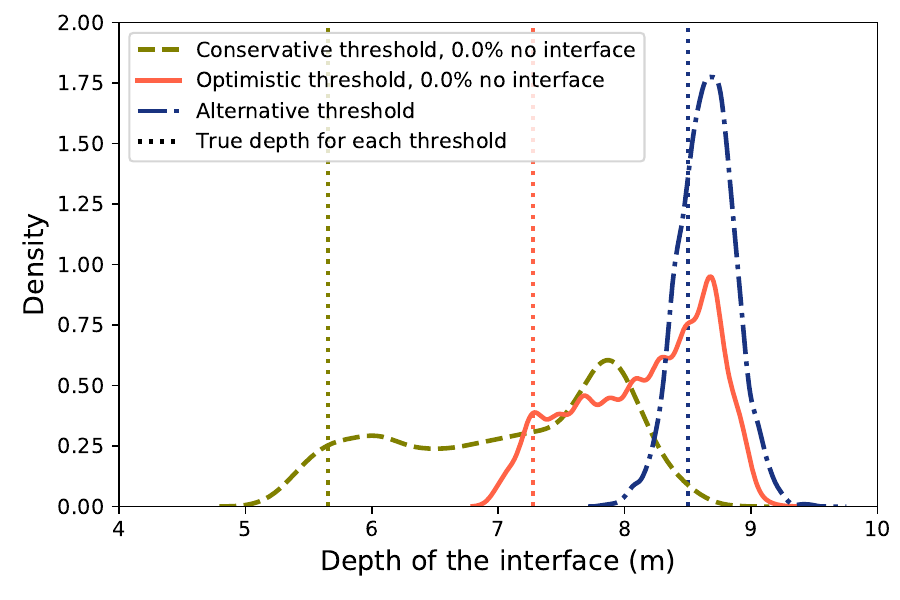}
	\caption{The posterior distribution of the depth of the interface according to the three thresholds.}
	\label{fig:benchmark_theshold_histogram}
\end{figure}

After projecting each model from the posterior to the depth of the interface according to the conservative, optimistic and alternative threshold defined in Section \ref{subsec:thresholds}, the posterior distribution of the depth of the interface is obtained. The true depth is derived from the true model, shown as dotted vertical lines in Figure \ref{fig:benchmark_theshold_histogram}, and differs for each definition of the interface. The conservative threshold yields the least certain depth estimation; the average discrepancy is 1.42 m, while this is 0.89 and 0.15 for the optimistic and alternative thresholds, respectively. Moreover, the standard deviation of the conservative threshold is 4.5 times the standard deviation of the alternative threshold, while this is only 2.5 for the optimistic case. In general, and especially for the optimistic and conservative threshold, the depth of the interface is overestimated compared to the true. This result is in contrast with earlier work with deterministic inversion, which generally underestimates the depths \citep{deleersnyder2024validation}. However, in this case, we should take into account that the prior distribution has an overabundance of sharper models than the true; sharper models tend to overestimate the depth. However, this does not affect the result about the width of the posterior. If the reliability of the depth estimation is an essential factor, the alternative threshold gives the most reliable result. 

\subsection{Effect of different sources of uncertainty}
\subsubsection{Uncertainty on flight altitude}

\begin{figure}
	\centering
	\includegraphics[width=\linewidth]{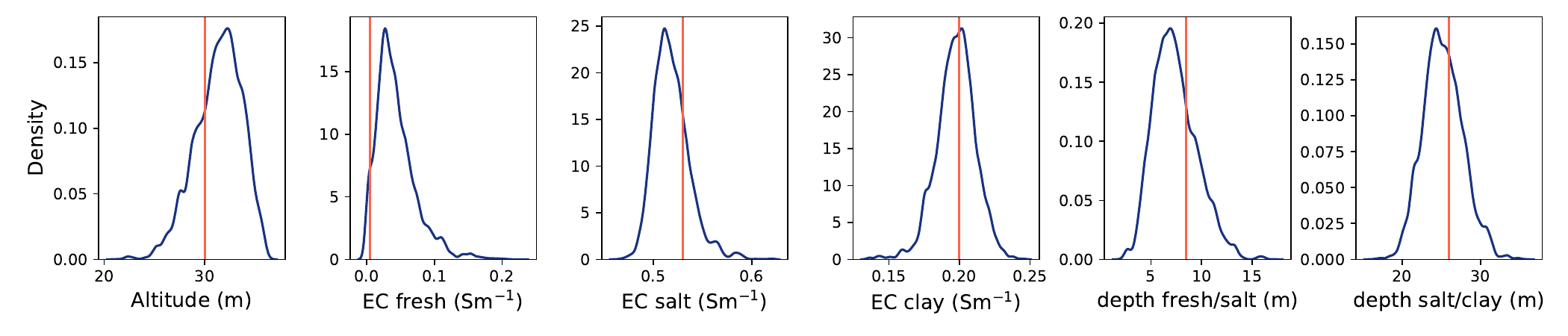}
	\includegraphics[width=\linewidth]{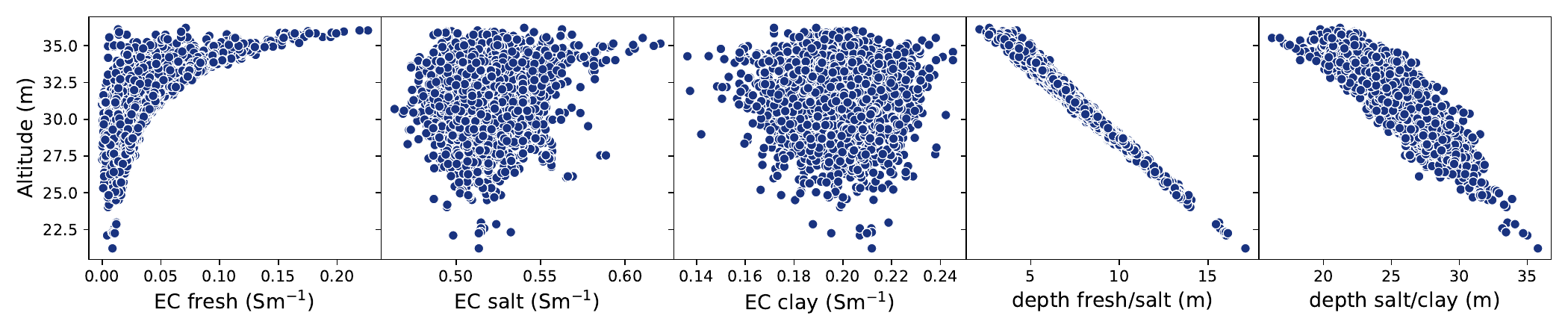}
	\caption{The posterior distribution for an uncertain flight altitude (2.5 m variance). The EC of the freshwater lens and the depths of the transitions are significantly affected.}
	\label{fig:uncertainty_altitude_correlations}
\end{figure}

In the previous section, the flight altitude was set at 30 m, and perfect knowledge was assumed. The synthetic data is identical in this section, but a measurement error in the altitude is assumed. The centre of the prior distribution is  30 m with a 2.5 m variance. The main results are shown in Figure \ref{fig:uncertainty_altitude_correlations}, namely the posterior distribution for each parameter except the sharpness because no significant effects were observed compared to the previous case. The correlation of the height parameter with the other parameters is also shown. Note that the prior was set to match the true flight altitude; however, the data indicates a slightly higher flight altitude is more likely. In addition, it is apparent that a much wider range of conductivities for the freshwater lens is possible. These are mainly possible at higher flight altitudes, while low ECs are possible at all sampled flight altitudes. The correlation between flight altitude and depth of the transitions was to be expected: models with deeper transitions correspond to lower altitudes. Based on these results, it is clear that the uncertainty of flight altitude has a large impact on the subsurface models and, consequently, the estimation of the depth of the interface.\\

\begin{figure}
	\centering
	\includegraphics[width=0.495\linewidth]{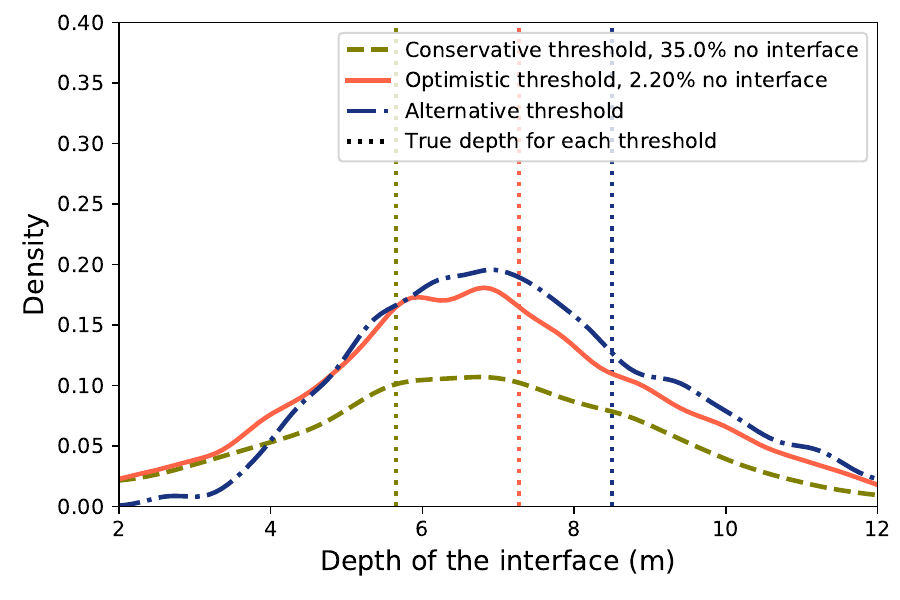}
	\caption{The posterior distribution of the depth of the interface according the three thresholds for an uncertain flight altitude (2.5 m variance).}
	\label{fig:uncertain_altitude_theshold_histogram}
\end{figure}

In Figure \ref{fig:uncertain_altitude_theshold_histogram}, the posterior of the depth is presented. A difference with Section \ref{subsec:benchmark} with certain flight altitude is that in 35\% of the cases, no interface is found with the conservative threshold because the EC of the freshwater lens was already larger than 0.05 Sm$^{-1}$. With the optimistic threshold, this is only 2.2\%, while the alternative threshold does not exhibit this problem. The optimistic threshold has, on average, the lowest discrepancy with the true depth.

\subsubsection{Effect of flight altitude}
The higher the altitude, the smaller the recorded magnetic field strength. 
With perfect knowledge of the altitude (no variance) and only a relative error (here, 3\% of the observed data), the depth of the interface is not affected (results not shown). In reality, however, one has to consider the signal-to-noise ratio and, therefore, include a noise floor (see Section REF). For high altitudes, more late-time gates will disappear below the noise floor and, thus, will no longer contribute to the likelihood.\\

\begin{figure}
	\centering
	\includegraphics[width=\linewidth]{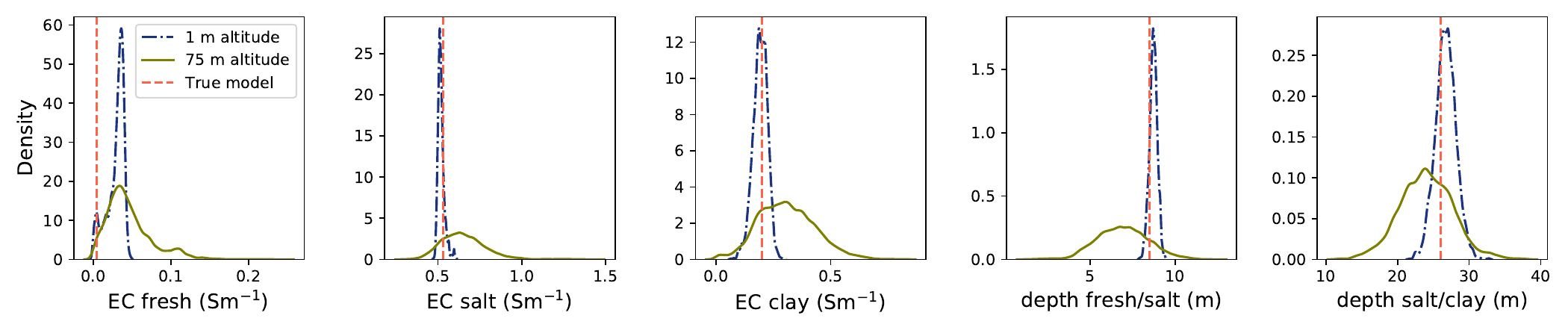}
	\caption{Effect of flight altitude on the estimation of the subsurface parameters.}
	\label{fig:effect_altitude_theshold_histogram}
\end{figure}

In Figure \ref{fig:effect_altitude_theshold_histogram}, we compare how accurately and certainly the parameters are recovered. The parameters at 1 m altitude are recovered with a much smaller variance than at 75 m altitude. Note again the bimodal character of the posterior distribution of the electrical conductivity of the freshwater. At 75 m altitude it is noticeable that there is a significant bias in the estimations of the depth.\\

\begin{figure}
	\centering
	\includegraphics[width=0.495\linewidth]{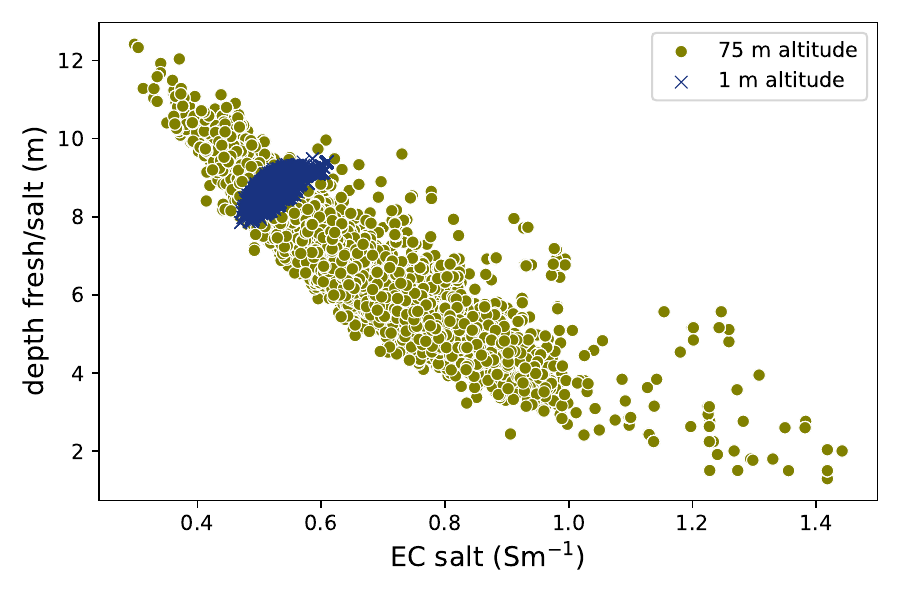}
	\caption{Effect of flight altitude on the covariances: a low altitude yields a smaller posterior space.}
	\label{fig:effect_altitude_covariances}
\end{figure}

Little changes in the correlations between parameters, except that a positive correlation was observed for the depths in terms of the EC of salt and clay. At higher altitudes, the correlation gradually became negative. In Figure \ref{fig:effect_altitude_covariances}, we illustrate this for the depth of the interface in terms of the EC of the saltwater lens. Interestingly, the posterior space at 1 m altitude forms a subspace of the posterior of the 75 m case. These effects are not observed if no noise floor is taken into account.

\begin{figure}
	\centering
	\includegraphics[width=0.495\linewidth]{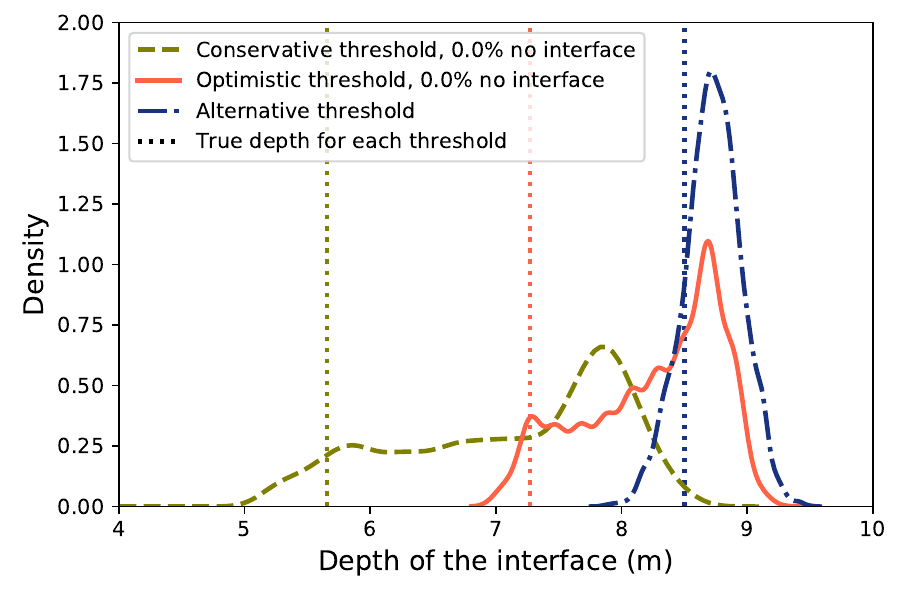}
	\includegraphics[width=0.495\linewidth]{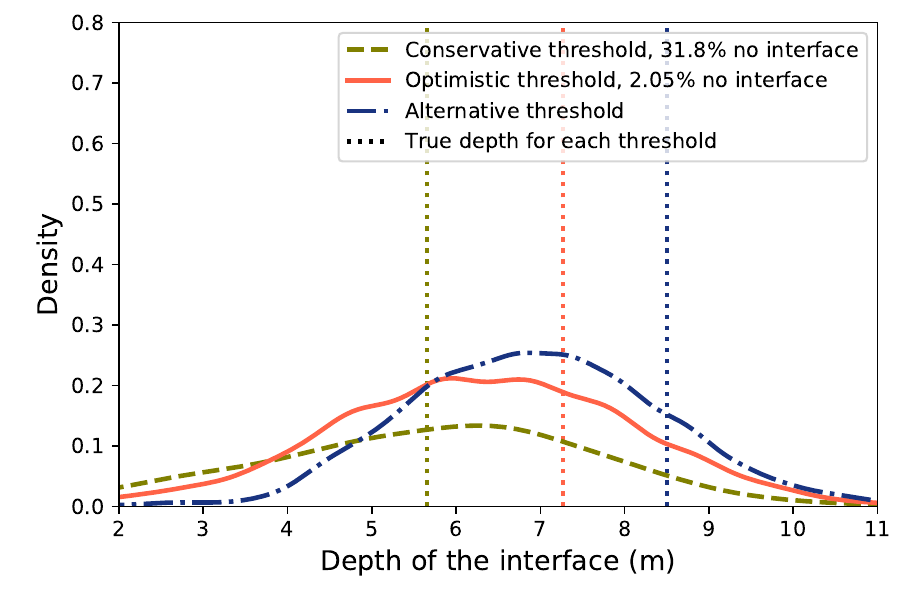}
	\caption{The posterior distribution of the depth of the interface according the three thresholds for different altitudes. \textbf{Left}: 1 m altitude, \textbf{right}: 75 m altitude. }
	\label{fig:effect_altitude_histogram_thresholds}
\end{figure}

For the posterior of the depth interface, according to the different thresholds in Figure \ref{fig:effect_altitude_histogram_thresholds}, we find a posterior at 1 m altitude similar to the case at 30 m altitude in Section \ref{subsec:benchmark}. In contrast, for high altitudes (and remember that we assume perfect knowledge of the altitude here), we find a similar result if we assume a variance of 2.5 at a height of 30 m.

\subsubsection{Uncertainty on the source: Pitch \& roll}
\begin{figure}
	\centering
	\includegraphics[width=0.495\linewidth]{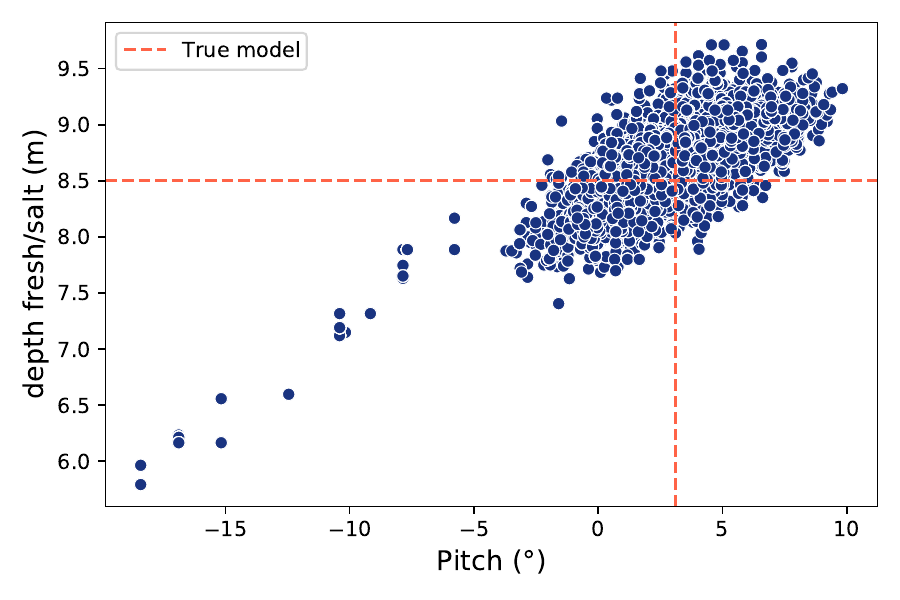}
	\includegraphics[width=0.495\linewidth]{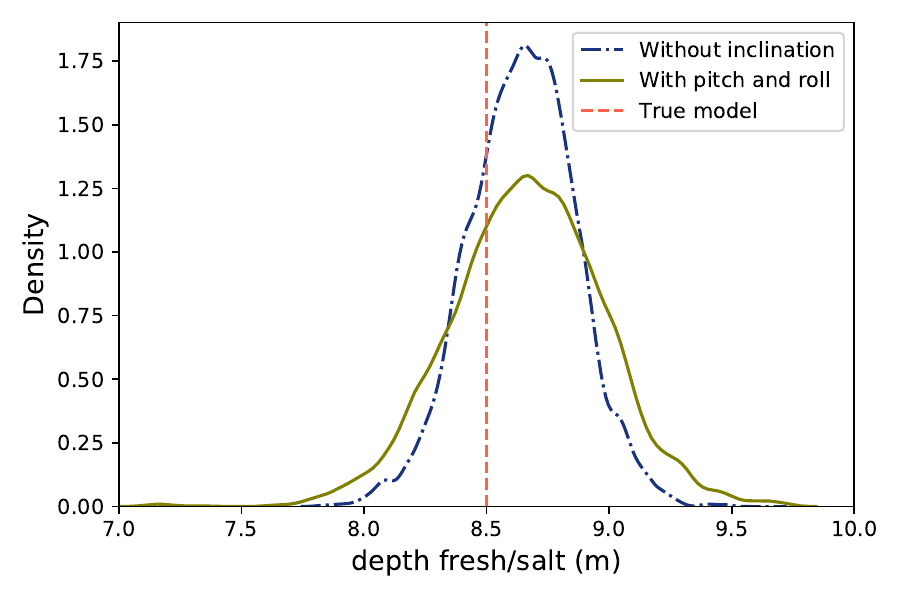}
	\caption{The effect of inclination of the transmitter-receiver on the posterior distribution.}
	\label{fig:effect_tilt}
\end{figure}

Airborne methods are subject to frame or bird motion. This section considers the differing transmitter and receiver heights and orientations caused by the various tilts. The true pitch and roll are 3.1 and -0.9, respectively, with 1.8 and 3.7 variance. Note that the McMC required a slightly longer burn-in time for this more complex case. The posterior distribution for the pitch and roll exactly follows the prior, meaning that the data itself cannot inform about the true tilt or roll. However, a correlation between the pitch and the depth of the transition from fresh to salt was observed (see Figure \ref{fig:effect_tilt}). Without an appropriate prior for the pitch, a pitch of less than -5° would have been explored much more, resulting in a significant deviation from the true depth. The few models do indicate the correlation, but do not contribute significantly to the density. In general, the density of the depth of the interface exhibits a slightly larger variance than without any inclination. The results indicate that the prior should be composed carefully and especially not ignored, which is a common practise in deterministic inversion. 

\subsubsection{Effect of salinity}
\label{subsec:salinity}
\begin{figure}
	\centering
	\includegraphics[width=0.495\linewidth]{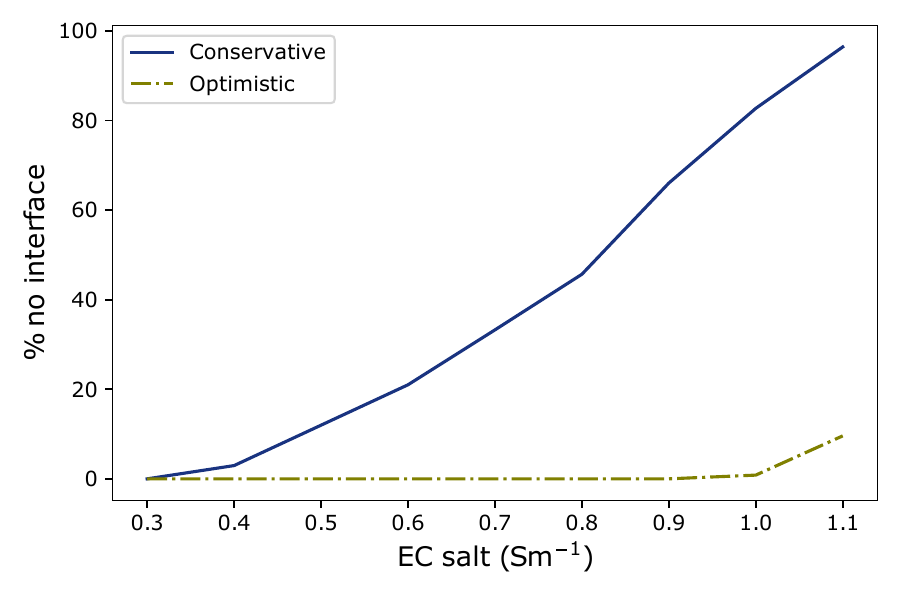}
	\includegraphics[width=0.495\linewidth]{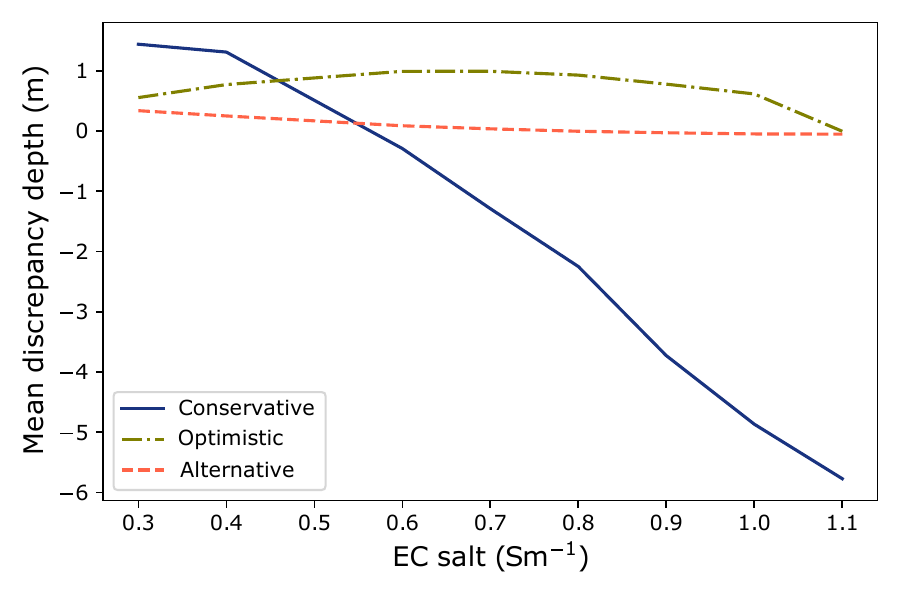}
	\caption{The effect of the electrical conductivity of the saltwater lens on the estimations of the interfaces' depth. \textbf{Left} shows the proportion of cases where no interface is found (only saltwater and clay). \textbf{Right} indicates the mean discrepancy from the true depth.}
	\label{fig:effect_salinity}
\end{figure}

As the salinity content in the study area varies (and potentially the empirical parameters in the petrophysical relation), it is essential to test the reliability of the threshold with respect to different electrical conductivities of the saltwater lens. A shift of 0.07 Sm$^{-1}$ for the centre of the prior with respect to the true is assumed (see Table \ref{tab:prior_benchmark}). The percentage at which no interface is observed increases with increasing salinity because (see Figure \ref{fig:effect_salinity}), for those models, the EC of the freshwater is above the threshold value. The deviation of the average depth of the posterior for each threshold with respect to the EC of the saltwater lens shows that the alternative threshold generates the best result for all conductivities. The optimistic threshold is much more stable than the conservative threshold.

\subsubsection{Less uncertainty on the sharpness}
\label{subsec:sharpness}
\begin{figure}
	\centering
	\includegraphics[width=0.495\linewidth]{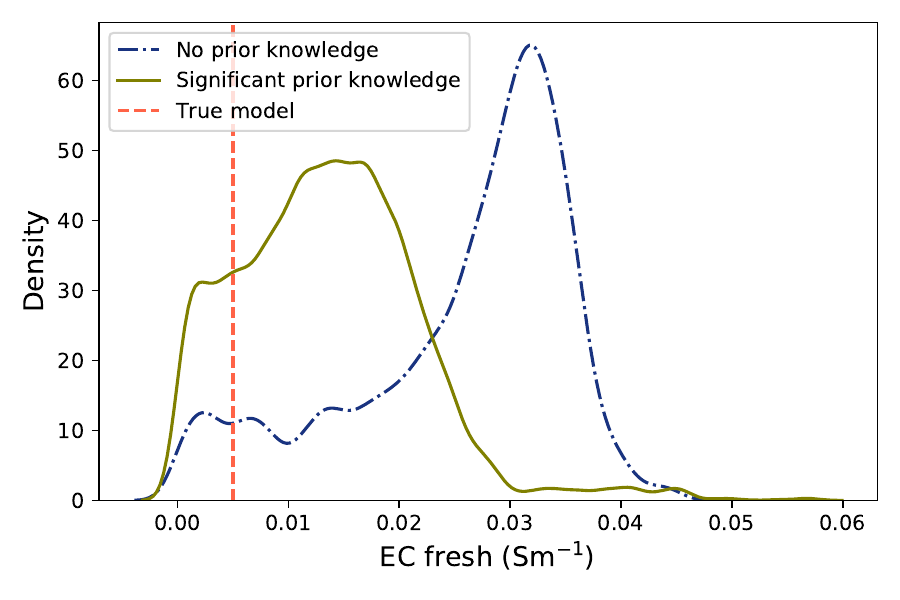}
	\includegraphics[width=0.495\linewidth]{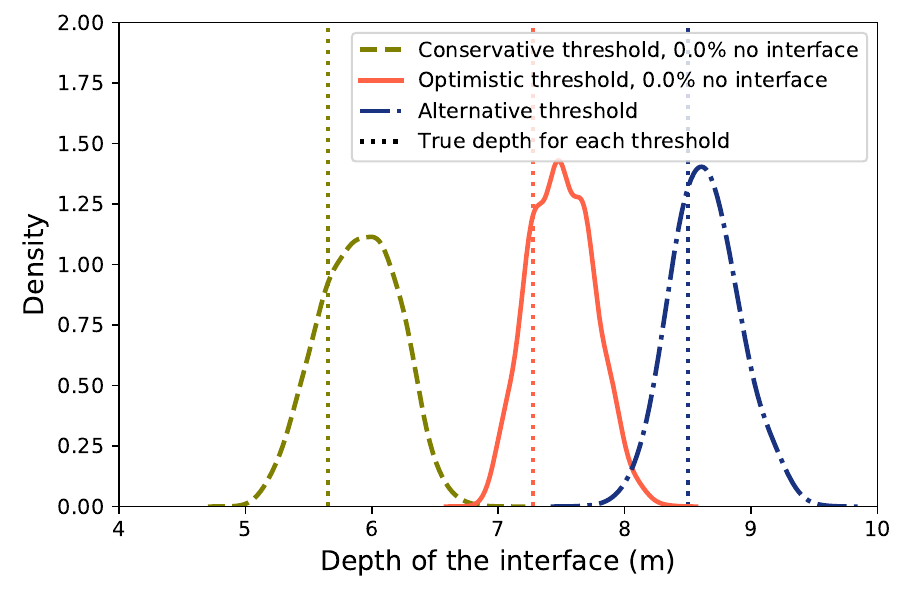}
	\caption{The effect of more prior information about the sharpness of the transition zone.}
	\label{fig:effect_sharpness}
\end{figure}

Suppose we have a good idea about the model's sharpness (e.g., via borehole loggings in the vicinity), and therefore can reduce the variance in the prior distribution of the sharpness. We assume the centre of the truncated normal distributions at 0.5 and 1 for the first and second transition, respectively, with both 1 as variance. Note that this prior knowledge can compensate for the inconsistency of 0.05 Sm$^{-1}$ as the centre of the prior for the EC of the fresh part by shifting towards true (see Figure \ref{fig:effect_sharpness}). The most significant difference is in the posterior of the depth of the interface in Figure \ref{fig:effect_sharpness}. Prior knowledge about the sharpness of the transitions has an enormous impact on the correct estimation of the depth for each threshold.

\subsubsection{FDEM data}
\begin{figure}
	\centering
	\includegraphics[width=\linewidth]{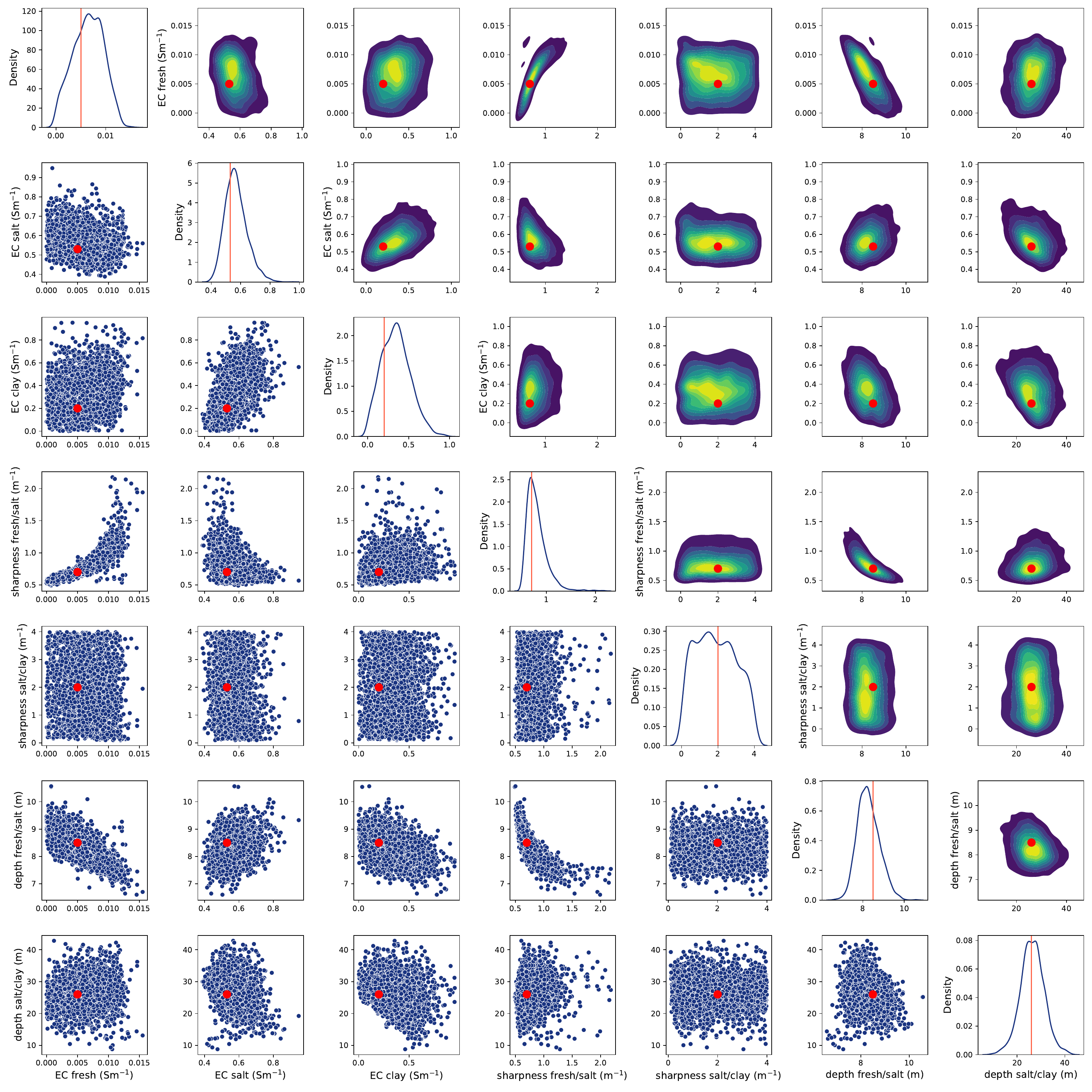}
	\caption{Prior distribution according to parametrization \eqref{eq:parameterization} reflecting the available prior information for the case in Section \ref{subsec:benchmark} \textit{with an FDEM system} and without altitude or inclination uncertainty.}
	\label{fig:FDEM}
\end{figure}

\begin{figure}
	\centering
	\includegraphics[width=0.495\linewidth]{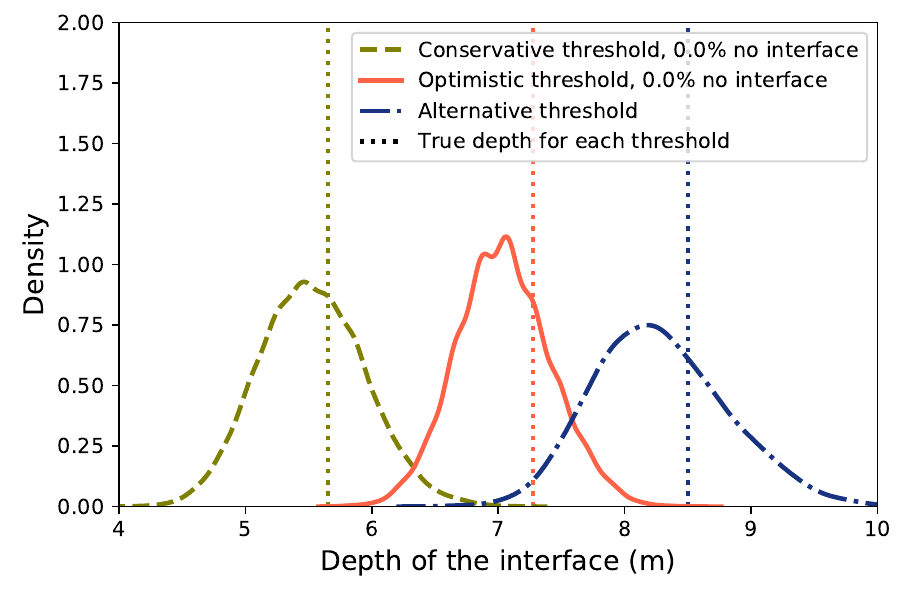}
	\caption{The posterior distribution of the depth of the interface according the three thresholds with an FDEM system. The posterior should be compared to the posterior in Figure \ref{fig:benchmark_theshold_histogram}.}
	\label{fig:effect_FDEM}
\end{figure}

This section examines how the choice of EM technology affects uncertainty. We choose a frequency-domain EM system, namely the RESOLVE system, which is known to have a better near-surface sensitivity. The system records only six data points: 5 horizontal coplanar coil orientations with 7.86 m at 382, 1~822,  7~970, 35~920, and 130~100 Hz, and one vertical coaxial coil orientation at 8.99 m at 3258 Hz. The prior is identical to Section \ref{subsec:benchmark} and Table \ref{tab:prior_benchmark}. The results are shown in Figure \ref{fig:FDEM}. The improvement in the freshwater lens's posterior distribution of the EC is immediately noticeable. The maximum of the distribution corresponds to the true conductivity. Note that the sharpness of the first transition is recovered very well, while almost no prior information was imposed on it. The transition from salt to clay is less well recovered. Figure \ref{fig:effect_FDEM}, showing the posterior distribution of the depth of the interface, should be compared with Figure \ref{fig:benchmark_theshold_histogram}. There is a significant improvement with an FDEM system. There is always a slight bias towards a too-shallow estimation of the interface, while with TDEM, this was just the opposite.

%

\section{Conclusion}
\label{sec:conclusion}
In this work, we have examined various factors that influence the accurate and reliable estimation of the depth of the fresh/saltwater transition with a time-domain AEM system. The findings can help to plan future AEM surveys, particularly those recommendations aimed at mapping the fresh/saltwater interface.\\

The findings are summarized as follows:
\begin{itemize}
	\item The choice of a threshold greatly impacts the reliability of the depth. The alternative threshold, the middle of the transition as a definition for the interface, is the most stable criterion for such an interface. The conservative (or straightforward) threshold was the least reliable. Additional thresholds can be easily tested in the manner described above.
	\item It pays to invest in more reliable altitude logging. Altitude uncertainty significantly increases the overall uncertainty and creates an extra bias in the posterior distribution. 
	\item The altitude affects reliability via the signal-to-noise ratio: the lower, the better. 
	\item Pitch and roll increase uncertainty but, unlike altitude uncertainty, do not introduce extra bias: the mean of the posterior remains unchanged.
	\item The salinity of the environment plays a role. The conservative threshold, especially, is prone to inaccurate estimations and is, therefore, strongly dependent on the bulk electrical conductivity of the saltwater lens. The alternative threshold is the most robust, and the optimistic threshold scores in between.
	\item If possible, prior knowledge of the expected sharpness should be included, for example through recording borehole logs, as it leads to a much better estimation of the depth of the interface.
	\item A frequency-domain system like the Resolve-system may be more suitable for salinity mapping than a time-domain SkyTEM system used in \citet{vlaanderentopsoil}. Further research should study the uncertainty due to specific issues related to such systems, such as drift. 
\end{itemize}

\section*{Acknowledgments}

The authors thank VMM (Flanders Environment Agency) for providing access to the boreholes.  The research leading to these results has received funding from the KU Leuven Postdoctoral Mandate PDMT2/23/065. The authors declare no conflicts of interest.

\bibliography{../../../literatuur/phd_sources}

\begin{thebibliography}{}

\bibitem[Ahmed et~al., 2024]{ahmed2024assessing}
Ahmed, A., Aigner, L., Michel, H., Deleersnyder, W., Dudal, D., Flores~Orozco,
  A., and Hermans, T. (2024).
\newblock Assessing and improving the robustness of bayesian evidential
  learning in one dimension for inverting time-domain electromagnetic data:
  Introducing a new threshold procedure.
\newblock {\em Water}, 16(7):1056.

\bibitem[Archie, 1942]{archie1942electrical}
Archie, G.~E. (1942).
\newblock The electrical resistivity log as an aid in determining some
  reservoir characteristics.
\newblock {\em Transactions of the AIME}, 146(01):54--62.

\bibitem[Auken et~al., 2009]{auken2009integrated}
Auken, E., Christiansen, A.~V., Westergaard, J.~H., Kirkegaard, C., Foged, N.,
  and Viezzoli, A. (2009).
\newblock An integrated processing scheme for high-resolution airborne
  electromagnetic surveys, the skytem system.
\newblock {\em Exploration Geophysics}, 40(2):184--192.

\bibitem[Bai et~al., 2021]{bai20211d}
Bai, P., Vignoli, G., and Hansen, T.~M. (2021).
\newblock 1{D} stochastic inversion of airborne time-domain electromagnetic
  data with realistic prior and accounting for the forward modeling error.
\newblock {\em Remote Sensing}, 13(19):3881.

\bibitem[Brodie and Reid, 2013]{brodie2013monte}
Brodie, R. and Reid, J. (2013).
\newblock Monte carlo inversion of skytem aem data from lake thetis, western
  australia.
\newblock {\em ASEG Extended Abstracts}, 2013(1):1--4.

\bibitem[Constable et~al., 1987]{constable1987occam}
Constable, S.~C., Parker, R.~L., and Constable, C.~G. (1987).
\newblock Occam’s inversion: A practical algorithm for generating smooth
  models from electromagnetic sounding data.
\newblock {\em Geophysics}, 52(3):289--300.

\bibitem[De~Breuck et~al., 1989]{debreuck1989}
De~Breuck, W., Beeuwsaert, E., and Vandenheede, J. (1989).
\newblock Diepte van het grensvlak tussen zoet en zout water in de freatische
  watervoerende laag van noordelijk {V}laanderen (1974-1975).
  {V}erziltingskaart {Z}eeuws-{V}laanderen.
\newblock Technical report, Universiteit Gent, Laboratorium voor Toegepaste
  Geologie en Hydrogeologie.

\bibitem[Deleersnyder, 2023]{deleersnyder2023improving}
Deleersnyder, W. (2023).
\newblock {\em Improving airborne time-domain electromagnetic imaging with
  applications to groundwater salinity mapping}.
\newblock PhD thesis.

\bibitem[Deleersnyder et~al., 2024]{deleersnyder2024validation}
Deleersnyder, W., Hermans, T., and Dudal, D. (2024).
\newblock Validation of the wavelet-based deterministic inversion ensemble
  approach with markov-chain monte carlo.
\newblock In {\em NSG2024 4th Conference on Airborne, Drone and Robotic
  Geophysics}, volume 2024. European Association of Geoscientists \& Engineers.

\bibitem[Deleersnyder et~al., 2021]{deleersnyder2021inversion}
Deleersnyder, W., Maveau, B., Hermans, T., and Dudal, D. (2021).
\newblock Inversion of electromagnetic induction data using a novel
  wavelet-based and scale-dependent regularization term.
\newblock {\em Geophysical Journal International}, 226(3):1715--1729.

\bibitem[Deleersnyder et~al., 2023]{deleersnyder2023flexible}
Deleersnyder, W., Maveau, B., Hermans, T., and Dudal, D. (2023).
\newblock Flexible quasi-2{D} inversion of time-domain {AEM} data, using a
  wavelet-based complexity measure.
\newblock {\em Geophysical Journal International}, 233(3):1847--1862.

\bibitem[Delsman et~al., 2019]{vlaanderentopsoil}
Delsman, J., van Baaren, E., Vermaas, T., Karaoulis, M., Bootsma, H., de~Louw,
  P., Pauw, P., Oude~Essink, G., Dabekaussen, W., Van~Camp, M., Walraevens, K.,
  Vandenbohede, A., Teilmann, R., and Thofte, S. (2019).
\newblock {TOPSOIL} {A}irborne {EM} kartering van zoet en zout grondwater in
  {V}laanderen.
\newblock Technical report, VMM.

\bibitem[Elshall et~al., 2020]{elshall2020groundwater}
Elshall, A.~S., Arik, A.~D., El-Kadi, A.~I., Pierce, S., Ye, M., Burnett,
  K.~M., Wada, C.~A., Bremer, L.~L., and Chun, G. (2020).
\newblock Groundwater sustainability: A review of the interactions between
  science and policy.
\newblock {\em Environmental Research Letters}, 15(9):093004.

\bibitem[Farquharson and Oldenburg, 1998]{farquharson1998non}
Farquharson, C.~G. and Oldenburg, D.~W. (1998).
\newblock Non-linear inversion using general measures of data misfit and model
  structure.
\newblock {\em Geophysical Journal International}, 134(1):213--227.

\bibitem[Gottschalk et~al., 2020]{gottschalk2020using}
Gottschalk, I., Knight, R., Asch, T., Abraham, J., and Cannia, J. (2020).
\newblock Using an airborne electromagnetic method to map saltwater intrusion
  in the northern salinas valley, california.
\newblock {\em Geophysics}, 85(4):B119--B131.

\bibitem[Hansen, 2021]{hansen2021efficient}
Hansen, T.~M. (2021).
\newblock Efficient probabilistic inversion using the rejection
  sampler—exemplified on airborne em data.
\newblock {\em Geophysical Journal International}, 224(1):543--557.

\bibitem[Hansen and Finlay, 2022]{hansen2022use}
Hansen, T.~M. and Finlay, C.~C. (2022).
\newblock Use of machine learning to estimate statistics of the posterior
  distribution in probabilistic inverse problems—an application to airborne
  em data.
\newblock {\em Journal of Geophysical Research: Solid Earth}, 127(11).

\bibitem[Hansen and Minsley, 2019]{hansen2019inversion}
Hansen, T.~M. and Minsley, B.~J. (2019).
\newblock Inversion of airborne em data with an explicit choice of prior model.
\newblock {\em Geophysical Journal International}, 218(2):1348--1366.

\bibitem[Hansen et~al., 2009]{hansen2009reducing}
Hansen, T.~M., Mosegaard, K., and Cordua, K.~S. (2009).
\newblock Reducing complexity of inverse problems using geostatistical priors.
\newblock In {\em International association of Mathematical geoscience (IAMG
  09)}.

\bibitem[Hunziker et~al., 2015]{hunziker2015electromagnetic}
Hunziker, J., Thorbecke, J., and Slob, E. (2015).
\newblock The electromagnetic response in a layered vertical transverse
  isotropic medium: A new look at an old problem.
\newblock {\em Geophysics}, 80(1):F1--F18.

\bibitem[King, 2022]{king2022airborne}
King, J.~A. (2022).
\newblock {\em Airborne electromagnetic mapping of coastal groundwater
  salinity: Quantifying uncertainty and investigating methodological
  improvements}.
\newblock PhD thesis, Utrecht University.

\bibitem[Klose et~al., 2022]{klose2022laterally}
Klose, T., Guillemoteau, J., Vignoli, G., and Tronicke, J. (2022).
\newblock Laterally constrained inversion ({LCI}) of multi-configuration {EMI}
  data with tunable sharpness.
\newblock {\em Journal of Applied Geophysics}, 196:104519.

\bibitem[Laloy and Vrugt, 2012]{laloy2012high}
Laloy, E. and Vrugt, J.~A. (2012).
\newblock High-dimensional posterior exploration of hydrologic models using
  multiple-try dream (zs) and high-performance computing.
\newblock {\em Water Resources Research}, 48(1).

\bibitem[Macnae and Milkereit, 2007]{macnae2007developments}
Macnae, J. and Milkereit, B. (2007).
\newblock Developments in broadband airborne electromagnetics in the past
  decade.
\newblock In {\em Proceedings of Exploration}, volume~7, pages 387--398.

\bibitem[Mikucki et~al., 2015]{mikucki2015deep}
Mikucki, J.~A., Auken, E., Tulaczyk, S., Virginia, R., Schamper, C.,
  S{\o}rensen, K., Doran, P., Dugan, H., and Foley, N. (2015).
\newblock Deep groundwater and potential subsurface habitats beneath an
  {A}ntarctic dry valley.
\newblock {\em Nature communications}, 6(1):1--9.

\bibitem[Oh and Byun, 2021]{oh2021bayesian}
Oh, S. and Byun, J. (2021).
\newblock Bayesian uncertainty estimation for deep learning inversion of
  electromagnetic data.
\newblock {\em IEEE Geoscience and Remote Sensing Letters}, 19:1--5.

\bibitem[Oude~Essink et~al., 2010]{oude2010effects}
Oude~Essink, G., Van~Baaren, E.~S., and De~Louw, P.~G. (2010).
\newblock Effects of climate change on coastal groundwater systems: A modeling
  study in the {N}etherlands.
\newblock {\em Water resources research}, 46(10).

\bibitem[Pedersen et~al., 2017]{pedersen2017mapping}
Pedersen, J.~B., Schaars, F.~W., Christiansen, A.~V., Foged, N., Schamper, C.,
  Rolf, H., and Auken, E. (2017).
\newblock Mapping the fresh-saltwater interface in the coastal zone using
  high-resolution airborne electromagnetics.
\newblock {\em First Break}, 35(8).

\bibitem[Pfaffhuber et~al., 2017]{pfaffhuber2017delineating}
Pfaffhuber, A.~A., Lysdahl, A.~O., S{\o}rmo, E., Skurdal, G.~H., Thomassen, T.,
  Ansch{\"u}tz, H., and Scheibz, J. (2017).
\newblock Delineating hazardous material without touching - {AEM} mapping of
  {N}orwegian alum shale.
\newblock {\em First Break}, 35(8).

\bibitem[Siemon et~al., 2019]{siemon2019automatic}
Siemon, B., van Baaren, E., Dabekaussen, W., Delsman, J., Dubelaar, W.,
  Karaoulis, M., and Steuer, A. (2019).
\newblock Automatic identification of fresh -- saline groundwater interfaces
  from airborne electromagnetic data in {Z}eeland, the {N}etherlands.
\newblock {\em Near Surface Geophysics}, 17(1):3--25.

\bibitem[Tarantola and Valette, 1982]{tarantola1982generalized}
Tarantola, A. and Valette, B. (1982).
\newblock Generalized nonlinear inverse problems solved using the least squares
  criterion.
\newblock {\em Reviews of Geophysics}, 20(2):219--232.

\bibitem[Taylor et~al., 2013]{taylor2013ground}
Taylor, R.~G., Scanlon, B., D{\"o}ll, P., Rodell, M., Van~Beek, R., Wada, Y.,
  Longuevergne, L., Leblanc, M., Famiglietti, J.~S., Edmunds, M., et~al.
  (2013).
\newblock Ground water and climate change.
\newblock {\em Nature climate change}, 3(4):322--329.

\bibitem[UNESCO, 2019]{UNESCO_2019}
UNESCO (2019).
\newblock {IHP-VIII}: Water security. responses to local, regional, and global
  challenges.
\newblock Website
  https://en.unesco.org/themes/water-security/hydrology/IHP-VIII-water-security.
\newblock Accessed on May 22th, 2023.

\bibitem[Vandenbohede, 2014]{vandenbohede2014quartair}
Vandenbohede, A. (2014).
\newblock Quartair van de {K}ustvlakte en polders van de {W}esterschelde
  [quaternary of the coastal plain and western schelde river].
\newblock {\em Watervoerende lagen en grondwater in Belgi{\"e}/Aquif{\`e}res et
  eaux souterraines en Belgique [Aquifers and groundwater in Belgium].
  Academia, Gent, Belgium}, pages 5--15.

\bibitem[Vandenbohede and Lebbe, 2012]{vandenbohede2012groundwater}
Vandenbohede, A. and Lebbe, L. (2012).
\newblock Groundwater chemistry patterns in the phreatic aquifer of the central
  {B}elgian coastal plain.
\newblock {\em Applied Geochemistry}, 27(1):22--36.

\bibitem[Viezzoli et~al., 2009]{viezzoli2009spatially}
Viezzoli, A., Auken, E., and Munday, T. (2009).
\newblock Spatially constrained inversion for quasi 3{D} modelling of airborne
  electromagnetic data – an application for environmental assessment in the
  {L}ower {M}urray {R}egion of {S}outh {A}ustralia.
\newblock {\em Exploration Geophysics}, 40(2):173--183.

\bibitem[Vignoli et~al., 2015]{vignoli2015sharp}
Vignoli, G., Fiandaca, G., Christiansen, A.~V., Kirkegaard, C., and Auken, E.
  (2015).
\newblock Sharp spatially constrained inversion with applications to transient
  electromagnetic data.
\newblock {\em Geophysical Prospecting}, 63(1):243--255.

\bibitem[Werthmüller, 2017]{werthmueller2017open}
Werthmüller, D. (2017).
\newblock An open-source full 3{D} electromagnetic modeler for 1{D} {VTI} media
  in {P}ython: empymod.
\newblock {\em {GEOPHYSICS}}, 82(6):WB9--WB19.

\end{thebibliography}

\end{document}